\documentclass[sigconf]{acmart}

\usepackage{booktabs} 
\usepackage{subfigure}

\setcopyright{none}

\acmDOI{}

\acmISBN{}

\acmConference[SuperCheck'21]{}{First International Symposium on Checkpointing for Supercomputing}{2021}
\acmYear{}

{\description \item[\fbox{To Do}]~\sl%
  \typeout{----------- Something left to be done here!
    ------------------}}%
{\enddescription}%

\begin{document}
\title{VELOC: VEry Low Overhead Checkpointing in the Age of Exascale}

\author{Bogdan Nicolae}
\affiliation{%
  \institution{Argonne National Laboratory}
  \country{USA}
}
\email{bnicolae@anl.gov}

\author{Adam Moody}
\affiliation{%
  \institution{Lawrence Livermore National Laboratory}
  \country{USA}
}
\email{moody20@llnl.gov}

\author{Gregory Kosinovsky}
\affiliation{%
  \institution{Lawrence Livermore National Laboratory}
  \country{USA}
}
\email{kosinovsky1@llnl.gov}

\author{Kathryn Mohror}
\affiliation{%
  \institution{Lawrence Livermore National Laboratory}
  \country{USA}
}
\email{mohror1@llnl.gov}

\author{Franck Cappello}
\affiliation{%
  \institution{Argonne National Laboratory}
  \country{USA}
}
\email{cappello@anl.gov}

\renewcommand{\shortauthors}{B. Nicolae et al.}

\begin{abstract}
  Checkpointing large amounts of related data concurrently to stable
  storage is a common I/O pattern of many HPC applications. However,
  such a pattern frequently leads to I/O bottlenecks that lead to poor
  scalability and performance. As modern HPC infrastructures continue
  to evolve, there is a growing gap between compute capacity vs.  I/O
  capabilities. Furthermore, the storage hierarchy is becoming
  increasingly heterogeneous: in addition to parallel file systems, it
  comprises burst buffers, key-value stores, deep memory hierarchies
  at node level, etc. In this context, state of art is insufficient to
  deal with the diversity of vendor APIs, performance and persistency
  characteristics. This extended abstract presents an overview of
  VeloC (Very Low Overhead Checkpointing System), a checkpointing
  runtime specifically design to address these challenges for the next
  generation Exascale HPC applications and systems. VeloC offers a
  simple API at user level, while employing an advanced multi-level
  resilience strategy that transparently optimizes the performance and
  scalability of checkpointing by leveraging heterogeneous storage.
\end{abstract}


\maketitle

\section{Introduction}
\label{sec:intro}

High performance computing (HPC) applications produce massive amounts
of checkpointing data during their runtime, which is often used for
\emph{defensive} purposes, i.e. to employ a \emph{checkpoint-restart}
resilience strategy in case of failures. In addition, the increasing
convergence between HPC, big data analytics and artificial
intelligence prompted many new scenarios for checkpointing, both
\emph{productive} (e.g., algorithms that revisit previous intermediate
results, coupling of workflow components, introspection to understand
the evolution of a computation or data), and \emph{administrative}
(e.g., out-of-core computations, co-scheduling of batch jobs and
on-demand jobs using suspend-resume).

Checkpointing in the context of HPC is a challenging I/O pattern,
because the data coming from a large number of distributed processes
typically is coordinated to form a globally consistent state. This
generates high write concurrency that overwhelms the I/O bandwidth of
the system, leading to large performance overheads and poor
scalability.  In the quest to reach Exascale, many architectural
trade-offs are necessary, including a high degree of parallelism and a
growing gap between the compute capacity and I/O capabilities, which
means less I/O bandwidth will be available per compute unit, thereby
making checkpointing even more challenging.

To avoid this issue, many applications have switched from writing
checkpoints directly to an external storage repository (e.g. a
parallel file system) to more advanced approaches, such as
\emph{multi-level checkpointing}. The idea is to use the faster and
less reliable local storage of the compute nodes (or that of
neighboring nodes) to implement ``lighter'' resilience levels that
hold the checkpoints. This enables applications to survive a majority
of failures without interacting with an external storage repository,
thereby conserving the scarce I/O bandwidth.

Despite promising potential, multi-level checkpointing as implemented
by state of art are not sufficient at Exascale and many challenges
remain. First, the storage stack is increasingly heterogeneous both
regarding compute nodes (deep memory hierarchies combined with local
storage) and external repositories (burst buffers, key-value stores,
parallel file systems). Therefore, it is increasingly difficult to
design optimal I/O and resilience strategies that can take advantage
of all storage options simultaneously. Second, in a quest to
differentiate from competition, vendors propose storage solutions with
different performance/resilience characteristics and custom
APIs. Therefore, it is important to solve the problem of
\emph{portability} for both I/O and resilience. Third, performance and
scalability requirements have prompted a transition from blocking
to asynchronous strategies (i.e., block the application only while
writing to the fastest level, while performing the rest of the
operations in the background). However, background operations may
compete with the application for resources, generating runtime
interference that is not well understood and needs to be minimized
through mitigation strategies.

This extended abstract proposes VELOC, a low overhead checkpointing
system that is part of the Exascale Computing Project (ECP) and aims
to build a production-ready solution that is specifically designed to
address the challenges mentioned above. For the rest of this paper, we
briefly introduce the main features of VELOC and discuss several
recent results.

\section{VELOC: An Overview}

\paragraph{Hidden Complexity of Heterogeneous Storage}
To address the challenge of complexity due to heterogeneous storage, we
propose a simple API that enables each application process to declare
``critical'' memory regions that need to be part of a global
checkpoint.  When a global checkpoint needs to be taken, all processes
call a collective checkpointing primitive that handles all details
transparently. Using this approach, users never have to worry what
types of storage are available and how to use them.  Furthermore,
the separation between fine-grained declarations of critical memory
regions and the actual checkpoint request opens several optimization
opportunities compared with writing the critical data directly to
a local storage device or external repository, such as: optimized
serialization, fine-grain allocation and movement of checkpoint chunks
between different types of storage based on memory layout and access
patterns, synergies between resilience strategies, etc. For example,
recent work enables VELOC to take advantage of heterogeneous node-local
storage to minimize the duration of blocking for asynchronous
flushing to an external repository. In this case, there are non-obvious
producer-consumer patterns that form under I/O concurrency, for which
using the fastest storage may be suboptimal~\cite{VeloCIPDPS19}.

\paragraph{Optimized Asynchronous Multi-Level Strategies}
We propose an advanced multi-level approach that is based on the idea
of leveraging idle resources in order to advance the asynchronous
checkpointing strategies in the background without causing significant
interference. To this end, we envision two possible complementary
approaches. First, if the behavior of the application is predictable
(which is the case of many iterative HPC applications that naturally
exhibit a repetitive behavior), then the background operations can be
scheduled in such way that they use different resources than those
needed by the application at a given moment. To this end, machine
learning techniques based on sequence-to-sequence models are a
promising tool in predicting the application
behavior~\cite{PredictionEuroPar19}.  Second, the background
operations can be scheduled such that they run with lower priority. In
this case, the operating system will reduce contention by giving the
application a large time slice at the expense of making the background
operations less predictable.  To this end, performance modeling using
micro-benchmarks focused on interference patterns can be used to
control the priority.

\paragraph{ML-Optimized Checkpoint Intervals}
The combination of asynchronous techniques that leverage heterogeneous
storage makes it very difficult if not impossible to determine an
optimal checkpoint interval analytically (due to sheer complexity) or
by simulation of the failure scenarios (due to a large number of
parameters that result in a massive number of failure
scenarios). Under such circumstances, a promising direction is the use
of machine learning to reduce the search space of the simulation-based
approaches.  Specifically, by sampling a subset of representative
failure scenarios, the aim is to train a ML model that is capable of
filling the missing gaps in the search space in order to predict the
optimal checkpoint interval with high accuracy. In this regard,
preliminary work~\cite{CkptInt-SC19} shows that models based on neural
networks can be particularly effective for this task, outperforming
other approaches such as random forest.

\begin{figure}
   \centering
   \includegraphics[width=0.7\linewidth]{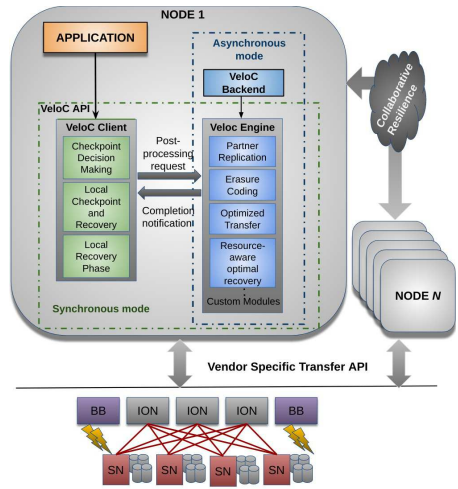}
   \caption{Architecture of VeloC (Very Low Overhead Checkpointing System)}
   \label{fig:arch}
\end{figure}

\paragraph{Flexibility through Modular Design}
We propose a modular design that encapsulates each I/O and resilience
strategy as an independent module that is part of a pipeline. Whenever
a checkpoint request is issued, the pipeline will trigger each module
one after another based on a pre-defined priority. Each module is
individually responsible to react to a checkpoint request and can do
so (or simply pass) based on its own internal state (e.g. optimal
checkpoint interval) and/or the outcome of the other modules invoked
earlier in the pipeline. Using this approach, a module can be
activated or deactivated at runtime as needed using a simple
switch. Furthermore, custom modules can be easily added in the
pipeline (e.g., conversion between output formats, compression,
integrity checks based on checksumming) and control when they are
triggered by customizing their priority with respect to the other
modules. The pipeline is managed by an engine that can run it either
\emph{synchronously} (in which case the engine is linked into
the application directly as a library) or \emph{asynchronously}
(in which case the engine lives is a separate process called the
active backend). This is illustrated in Figure~\ref{fig:arch}.

\section{Productive Checkpointing: The Case for Deep Learning}

As mentioned in Section~\ref{sec:intro}, productive use case of
checkpointing make it a valuable tool in the design of new algorithms
and approaches that revisit previous states. At the intersection of
HPC and deep learning, systematic approaches such as guided
\emph{model discovery} and \emph{sensitivity analysis} illustrate the
need to \emph{capture} intermediate snapshots of the DNN model in
order to study its evolution in time and potentially \emph{reuse} it
later.

For example, outlier detection is critical for evaluating, curating,
and using biomedical data sets used to train DNN models. In this case,
approaches such as~\cite{MLHPC20} build deep learning ensembles and workflows
to construct a system for automatically identifying data subsets that have a
large impact on the trained models. Specifically, the goal is to evaluate
many training variations with and without considering subsets of the
training samples. However, these training variations need not be trained
independently from scratch, they can share a common training path
up until a point when they begin to diverge. Therefore, the model
can be checkpointed and replicated in order to be able to branch off
in different directions, which greatly speeds up the exploration.

Our work on DeepFreeze~\cite{DeepFreeze20} illustrates DNN
checkpointing techniques based on the idea of augmenting the execution
graph with fine-grain tensor copy operations, which can run in
parallel with gradient computations and weight updates involving
different layers during the back-propagation. Using this approach, a
full checkpoint of the DNN model can be produced in-memory or on local
storage with minimal impact on the learning performance, which can
then be again transferred asynchronously to the memory of a different
GPU and/or remotely to a different compute node.

This can be further improved with techniques such as
DeepClone~\cite{DeepClone20}: the augmentation techniques for the
execution graph during the back-propagation can be extended with
additional techniques such as zero-copy transfers of tensors and
optimized reconstruction that efficiently replicates a DNN model in
the memory of remote nodes without involving stable storage. Furthermore
such techniques can take advantage of already existing replicas
that are naturally produced by large-scale data-parallel training
techniques.

Such considerations have inspired new data models such as \emph{data
states}~\cite{DataStates20}, which are intermediate snapshots of
datasets (e.g., DNN models) that can be either captured or cloned
asynchronously at scale, while making them discoverable and accessible
a lineage, making it easy to navigate through their evolution and/or
search for interesting snapshots that can be reused.

\section{Exascale Ecosystem}

VELOC is developed as part of the Exascale Computing Project (ECP) and is
designed as a production-ready multi-level asynchronous checkpointing
solution based on the features mentioned above. Currently, it serves several
ECP applications, including HACC, LatticeQCD and EXAALT.

It is in use and regularly tested on several pre-Exascale testbeds,
including Theta (4392 Intel Xeon Phi KNL nodes, peak: 11.69 PFLOPS),
Summit (9126 POWER9 CPUs, 27648 NVIDIA Tesla V100 GPUs, peak: 200
PFLOPS peak), Sierra (architecture similar to Summit, peak: 135
PFLOPS).  Other platforms where VELOC is currently being evaluated
include Frontera (8008 Intel Xeon nodes, peak: 38.7 PFLOPS) and Fugaku
(158976 ARM64 nodes, peak: 442 PFLOPS).

A recent run on Summit at full scale for the HACC application achieved
an I/O throughput of up to 224 TB/s for writing local in-memory
checkpoints in a blocking fashion, while generating a negligible
runtime overhead for flushing the local checkpoints to a Lustre
parallel file system in the background.

We aim to integrate VELOC with alternative external storage
repositories that complement parallel file systems. Notably, a recent
effort has targeted DAOS, a scalable object storage system developed
by Intel. To this end, we developed an experimental module that
leverages an optimized low-level put/get API for key-value pairs.

\section*{Acknowledgments}
This research was supported by the Exascale Computing
Project (ECP), Project Number: 17-SC-20-SC, a collaborative
effort of two DOE organizations – the Office of Science and
the National Nuclear Security Administration, responsible for
the planning and preparation of a capable exascale ecosystem,
including software, applications, hardware, advanced system
engineering and early testbed platforms, to support the nation’s
exascale computing imperative. This material was based upon
work supported by the U.S. Department of Energy, Office of
Science, under contract DE-AC02-06CH11357. This research
used resources of the Argonne Leadership Computing Facility,
which is a DOE Office of Science User Facility supported
under Contract DE-AC02-06CH11357. This work was performed under
the auspices of the U.S. Department of Energy
by Lawrence Livermore National Laboratory under Contract
DE-AC52-07NA27344.

\bibliographystyle{ACM-Reference-Format}
\bibliography{main}

\end{document}